\documentclass[aps,prd,notitlepage,amssymb,amsmath,floatfix,nofootinbib,superscriptaddress]{revtex4-1}
\usepackage{amsmath,graphicx,epsfig,amssymb,dsfont,mathtools}
\usepackage[usenames]{color}

\allowdisplaybreaks[4]

\begin{document}
\title{Study of lepton flavor universality and angular distributions in $D\to K^*_{J} (\to K\pi)\ell\nu_{\ell}$}
\author{Fei Huang}
\affiliation{School of Physics and Technology, University of Jinan, 250022, Jinan, China,}
\author{Ji Xu}
\email{Email:xuji\_phy@zzu.edu.cn}
\affiliation{School of Physics and Microelectronics, Zhengzhou University, Zhengzhou, Henan 450001, China}

\begin{abstract}
The decay of the D  meson into multibody final states is a complex process that provides valuable insights into the fundamental interactions within the Standard Model of particle physics. This study focuses on the decay cascade $D^{+}\to  K^{*}_{J} \ell^{+}\nu \to  K^{\pm}\pi^{\mp}  \ell^{+}\nu$  where the $K^*_J$ resonance encompasses the  $K^*(892),K^{*}(1410),K^{*}_0(1430)$ states. We employ the helicity amplitude technique to derive the angular distributions for the decay chain, enabling the extraction of one-dimensional and two-dimensional distributions. Utilizing form factors for the  $D\to K^*$  transition derived from the quark model, we calculate the differential and integrated partial decay widths, explicitly considering the electron and muon masses. 
 Decay branching fractions are calculated, the ratios of the branching fractions are found to be $\frac{\mathcal{B}r(D^{+}\to K^{*}(892)(\to K ^{-}\pi^{+}) \, \mu^{+}\nu_{\mu})}{\mathcal{B}r(D^{+}\to K^{*}(892)(\to K ^{-}\pi^{+}) \, e^{+}\nu_{e})} = 0.975$, $\frac{\mathcal{B}r(D^{+}\to K^{*}(1410)(\to K ^{-}\pi^{+}) \, \mu^{+}\nu_{\mu})}{\mathcal{B}r(D^{+}\to K^{*}(1410)(\to K ^{-}\pi^{+}) \, e^{+}\nu_{e})} = 0.714$ and $\frac{\mathcal{B}r(D^{+}\to K^{*}_{0}(1430)(\to K ^{+}\pi^{-}) \, \mu^{+}\nu_{\mu})}{\mathcal{B}r(D^{+}\to K^{*}_{0}(1430)(\to K ^{+}\pi^{-}) \, e^{+}\nu_{e})} = 0.774$. Results in this work will serve a calibration for the study of  $c \to s $ in $D$ meson decays in future and provide useful information towards the understanding of the properties of the $K^{*}$ meson, as well as $K \pi$ system.
\end{abstract}
\maketitle

%

\section{Introduction}
Research on the semileptonic  decay modes of heavy mesons offers an ideal laboratory to investigate both weak and strong interactions, in which the valuable information to extract the Standard Model (SM) parameters can be provided. A high precision study of these decay modes also presents an indirect pathway for new physics (NP). In the SM, the charged leptons (electron, muon, and $\tau$ lepton) share identical coupling to the gauge bosons, resulting in a similar behavior, up to the kinematic differences attributed to their different masses. This is commonly known as lepton flavor universality (LFU) \cite{ParticleDataGroup:2022pth}.

Rare $b$ hadron decays provide an excellent environment to test LFU. The semileptonic partial decay width is connected to the product of the hadronic form factor that describing the strong interaction in the initial and final states, and the Cabibbo-Kobayashi-Maskawa (CKM) matrix elements parametrizing the mixing between the quark flavors in the weak interaction. Detailed studies of the semileptonic decays allow for measurements of hadronic form factors, which are pivotal for calibrating the involved theoretical calculations \cite{Cheng:2003sm,Xing:2022uqu,Shen:2021yhe,Gao:2019lta,Lu:2022kos,Wang:2015paa,Meissner:2013pba}. The LHCb collaboration has reported several hints for lepton flavor universality violation (LFUV) in $B^+ \to K^+ \ell^+ \ell^-$ \cite{LHCb:2014vgu,LHCb:2019hip}, $B^0 \to K^*(892)^0 \, \ell^+ \ell^-$ \cite{LHCb:2017avl}, and $B^+ \to K^*(892)^+ \ell^+ \ell^-$ \cite{LHCb:2021lvy}. Although the assertion of potential LFUV in $B$ meson decays has diminished in the latest LHC results \cite{LHCb:2022qnv,LHCb:2022vje,CMS:2024syx,Smith:2024xgo}, the interest in these processes remains strong \cite{Ciuchini:2021smi,SinghChundawat:2022ldm,Ladisa:2022vmh,Dasgupta:2023zrh,Allanach:2024jls,Zaki:2023mcw,Farooq:2024owx,Alok:2024cyq,Wang:2014lda}.

On the other hand, the semileptonic $D$ decays are also of interest and importance. Accurate measurements of this kind of processes can serve as useful guidelines to semileptonic $B$ decays \cite{Meissner:2013hya,BaBar:2010vmf}. High statistics in $D^+ \to K^- \pi^+ e^+ \nu_e$ allows investigating the properties of $\bar K^*(892)^0$ meson. Furthermore, the semileptonic $D$ decay is a necessary ingredient in certain factorization approach for nonleptonic $D$ decay \cite{Escribano:2023zjx}. In the SM, semileptonic $D$ decays also offer an excellent opportunity to test LFU \cite{Li:2021iwf}. All of these advantages have attracted a great deal of attention to this kind of semileptonic decays, with the relative simplicity of theoretical description and the large branching fraction.

Due to the short lifetime of $K^{*}_{J}$ in the semileptonic decay $D^{+}\to  K^{*}_{J} \ell^{+}\nu$, the $ K^*_J$ resonance cannot be directly detected by experiments and must be reconstructed through the two or three pseudo-scalars $\pi/K$ final state. Therefore these decay modes are at least four-body processes. The decay channel $D^+ \to K^- \pi^+ e^+ \nu_e$ has been analyzed by FOCUS \cite{FOCUS:2002xsy}, CLEO \cite{CLEO:2010enr}, BABAR \cite{BaBar:2010vmf}, and BESIII \cite{BESIII:2015hty} collaborations. Although the structure of $K \pi$ system is dominated by the $\bar K^*(892)^0$, the  contributions of $\bar K^*(1410)^0$ and $\bar K^*_0(1430)^0$ is also considered. BESIII collaboration  reported that the contribution from $\bar K^*(1410)^0$ is minimal, while the contribution of the S-wave (the $\bar K^*_0(1430)^0$ and the nonresonant part) is observed with a significance far larger than $10\sigma$ \cite{BESIII:2015hty}. The contribution from the $\bar K^*(1680)^0$ can be ignored because it is suppressed by the small phase space available. Additionally, studies of the isospin-symmetric mode $D^0 \to \bar K^0 \pi^- e^+ \nu_e$ will provide complementary information on the $\bar K \pi$ system \cite{BESIII:2018jjm,BESIII:2024jlj,FOCUS:2004zbs,CLEO:2005cuk}. A ratio of branching fractions obtained in \cite{BESIII:2024jlj} is $\frac{\mathcal{B}r(D^{0}\to K^{*}(892)^- \, \mu^{+}\nu_{\mu})}{\mathcal{B}r(D^{0}\to K^{*}(892)^- \, e^{+}\nu_{e})} = 0.96 \pm 0.08$, in agreement with LFU.  Furthermore, we prefer to study the system of D meson decays with resonances to elucidate why  the contribution of $\bar K^*(1430)^0$  is larger than $\bar K^*(1410)^0$ and to test the LFU of these decays. 

The purpose of this work is multifold. We will first give the details in the helicity amplitude approach to derive the information about pertinent angular distributions of semileptonic $D^+$ decay. Secondly we calculate the differential and integrated partial decay widths based on the $D\to K^*$ form factors from quark model, with the electron and muon mass explicitly included. The branching fractions of $D^{+}\to  K^{*}_{J} \ell^{+}\nu$ are presented, our results align with the existing experimental data, indicating the flexibility of the form factors proposed in \cite{Cheng:2003sm}. Besides, we will derive the ratios of branching fractions with difference resonances of $K^{*}_{J}$, which might be useful in ongoing and future tests of lepton universality. Throughout this work we will use $K^*_J$ to abbreviate the vector meson $K^*(892)$, $K^*(1410)$ and the scalar meson $K^*_0(1430)$.

This paper is organized in the following way. In Sec.\,\ref{Fandad}, we give the theoretical framework for the semileptonic $D^+$ decay. The helicity amplitude is adopted to derive the angular distributions. The numerical results for $D^{+}\to  K^{*}_{J} \ell^{+}\nu$ is detailed in Sec.\,\ref{Panda}, followed by the conclusions in Sec.\,\ref{Conclusions}. Additional details of the coefficients that appeared in angular distribution can be found in the appendix.

\section{Framework and angular distribution}
\label{Fandad}
\begin{figure}[htbp!]
  \centering
  \includegraphics[width=0.7\linewidth]{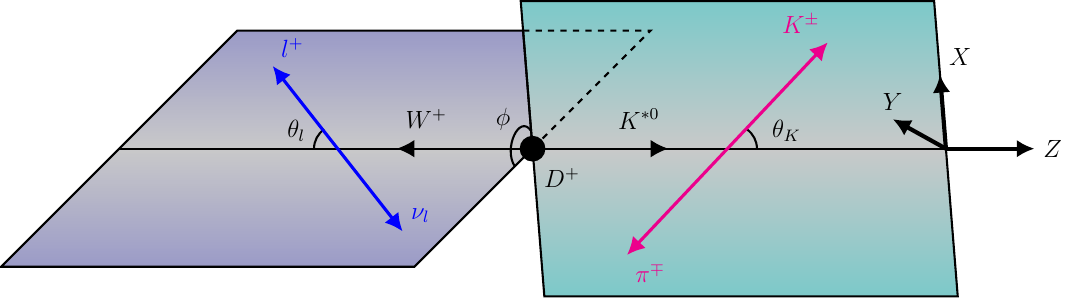}
 \caption{ The kinematics for the decay process of $D^{+}\to K_{J}^{*} (\to K\pi)\ell^{+}\nu_{\ell}$. In the $D$ meson rest frame, the momenta of $K_{J}^{*}$ is along the z-axis. The $\phi$ is the azimuthal angle between the $K_{J}^{*}$ and leptonic decay cascade decay planes. The $\theta_{K}$($\theta_{\ell}$) is defined as the angle between positive (negative) z-axis and momentum of $K (\ell$).   } \label{decay}
\end{figure}

The decay kinematics for $D^{+}\to K_{J}^{*}(\to K\pi)\ell^{+}\nu_{\ell}$ are illustrated in Figure \ref{decay}. There is no additional hadron in the final state, all the requisite  information for distinguishing the different hadronic angular momentum components can be derived from correlations between the leptonic and hadronic systems. In this work, we adopt the following convention. In the rest frame of the $D$ meson, the momentum of $K_{J}^{*}$ is directed along the z-axis. $\phi$ is the azimuthal angle between the planes defined by the hadronic and leptonic decays. $\theta_{K}$ is the polar angle of  the rest frame of $K_{J}^{*}$ and  $\theta_{\ell}$ is that of $\ell^{+}$ in the $W^{+}$ rest frame. In this section, we employ the helicity amplitude technique to analyze the multi-body decays of $D$ mesons involving three distinct $K_{J}^{*}$ resonances.
\subsection{Helicity Amplitude}
The effective electro-weak Hamiltonian governing the semileptonic decays of charmed mesons can be expressed as
\begin{equation}
\mathcal{H}_{\rm{eff}}=\frac{G_{F}}{\sqrt{2}}V_{cs}\big(\bar{s}\gamma^{\mu}(1-\gamma^{5})c\big)\big(\bar{\nu}_{\ell}\gamma_{\mu}(1-\gamma_{5})\ell)\big) \,,
\end{equation}
where $G_{F}$ is the Fermi constant and $V_{cs}$ is the CKM matrix element. $\ell$ denotes two types of leptons with flavor $e$ or $\mu$. According to the effective Hamiltonian, the relation between the metric tensor and polarization vector is $g_{\mu\nu}=-\sum_{(\lambda=0,\pm1,t)}\epsilon_{\mu}^{*}(\lambda)\epsilon_{\nu}^{*}(\lambda)+\frac{q_{\mu}q_{\nu}}{q^{2}}$. The amplitude for this four-body decay can be divided into a Lorentz-invariant hadronic part and a leptonic matrix element,
\begin{align}
\mathcal{M}(D\to K^{*}_{J}(\to K \pi)\ell\nu)&=\frac{G_{F}}{\sqrt{2}}V_{cs}\langle K\pi | \bar{s}\gamma^{\mu}(1-\gamma_{5})c|D\rangle\times\langle\ell\nu_{\ell}|\bar{\nu_{\ell}}\gamma^{\nu}(1-\gamma_{5})\ell |0\rangle g_{\mu\nu} \\ \nonumber
&=\frac{G_{F}}{\sqrt{2}}V_{cs}(-\sum_{\lambda}H \cdot \epsilon^{*}(\lambda)\times L \cdot \epsilon(\lambda)+H \cdot \epsilon^{*}(t)\times L \cdot \epsilon(t)) \,.
\end{align}
Here the hadronic and leptonic parts are given by $H^{\mu}=\langle K \pi |\bar{s}\gamma^{\mu}(1-\gamma_{5})c|D\rangle$ and $L^{\nu}=\langle \ell\nu_{\ell}|\bar{\nu_{\ell}}\gamma^{\nu}(1-\gamma_{5})\ell |0\rangle$, respectively. Besides, $\epsilon$ denotes the polarization of $W$ boson with helicity state $\lambda$ and $t$. $L^{\nu}$ represents the matrix element of $V-A$ current between the vacuum and final states of lepton and neutrino.

Utilizing the particles completeness states of the intermediate mesons
\begin{align}
\sum_{J}\int\frac{d^4 p^2_{K^{*}_{J}}}{(2\pi)^4}\frac{i}{p_{2}^{2}-m_{K^{*}_{J}}^2+im_{K^{*}_{J}}\Gamma_{K^{*}_{J}}}\times |K^{*}_{J} \rangle \langle K^{*}_{J}|=1 \;,
\end{align}
the hadronic part $H^{\mu}$  can be further calculated as
\begin{align}
H^{\mu}_{J,\lambda_{W}}=&\sum_{s_{K^{*}_{J}}}\int \frac{d^{4} p_{K^{*}_{J}} }{(2\pi)^4}\frac{i}{p_{2}^{2}-m_{K^{*}_{J}}^2+im_{K^{*}_{J}}\Gamma_{K^{*}_{J}}}\times \langle K(p_{k},s_{K}) \pi(p_{\pi})|K^{*}_{J}\rangle \\ \nonumber
&\times \langle K^{*}_{J}(p_{ K^{*}_{J}},s_{K^{*}_{J}})|\bar{s}\gamma^{\mu}(1-\gamma_{5})c| D(p_{D},s_{D})\rangle \,.
\end{align}
Here, $p_{K^{*}_{J}}$, $m_{K^{*}_{J}}$ and $\Gamma_{K^{*}_{J}}$ are the four-momentum, mass and decay width of intermediate state $K^{*}_{J}$, respectively.  $p_{D}$ is the momentum of the  initial state D meson. $\lambda_{W}$ is the helicity of off-shell W boson.

The transition matrix elements for the decay of $D$ mesons to the vector mesons $K^{*}_{892,1410}$ are parameterized in terms of the associated hadronic form factors
\begin{align}
\langle K^{*}_{J}|\bar{s}\gamma^{\mu}\gamma_{5}c|D\rangle =&
-\frac{2iA(q^2)}{m_{D}-m_{K^{*}_{J}}}\epsilon^{\mu\nu\rho\sigma}(\epsilon^{*}_{K^{*}_{J}})_{\nu}(p_{D})_{\rho}(p_{K^{*}_{J}})_{\sigma} \,,\\
\langle K^{*}_{J}|\bar{s}\gamma^{\mu}c|D\rangle =&-2m_{K^{*}_{J}}V_{0}(q^2)\frac{\epsilon_{K^{*}_{J}}\cdot q}{q^2}q^{\mu}-(m_{D}-m_{K^{*}_{J}})V_{1}(q^{2})\left[\epsilon^{*}_{K^{*}_{J}}-\frac{\epsilon_{K^{*}_{J}}\cdot q}{q^2}\right] \\ \nonumber
&+V_{2}\frac{\epsilon_{K^{*}_{J}}\cdot q}{m_{D}-m_{K^{*}_{J}}}\left[(p_{D}+p_{K^{*}_{J}})^{\mu}-\frac{m_{D}^{2}-m_{K}^{2}}{q^2}q^{\mu}\right] \,,
\end{align}
where $m_{D}$ and $m_{K^{*}_{J}}$ are the mass of initial and final states, respectively. $\epsilon^{\mu\nu\rho\sigma}$ is the antisymmetric tensor. $q=p_{D}-p_{K^{*}_{J}}$ is  the momentum transfer. $V_{i}(q^2)(i=0,1,2)$ and $A(q^2)$ are nonperturbative form factors. The matrix element $H_{\lambda_{W}}\equiv \langle K^{*}_{J}|\bar{s}\gamma^{\mu}\gamma_{5}c|D\rangle$ can be calculated as
\begin{subequations}
\begin{align}
H_{+} =&(m_{D}-m_{K^{*}_{J}})V_{1}-\frac{A\sqrt{\lambda(m_{D}^{2},m_{K^{*}_{J}}^2,q^2)}}{m_{D}-m_{K^{*}_{J}}} \,,\\
H_{-} =&(m_{D}-m_{K^{*}_{J}})V_{1}+\frac{A\sqrt{\lambda(m_{D}^{2},m_{K^{*}_{J}}^2,q^2)}}{m_{D}-m_{K^{*}_{J}}} \,,\\
H_{0} =&\frac{-1}{2m_{K^{*}_{J}}\sqrt{q^2}}\left[(m_{D}^2-m_{K^{*}_{J}}^2-q^2)(m_{D}-m_{K^{*}_{J}})
V_{1}-\frac{\lambda(m_{D}^{2},m_{K^{*}_{J}}^2,q^2)}{m_{D}-m_{K^{*}_{J}}}V_{2}\right] \,,\\
H_{t} =&-\frac{\sqrt{\lambda(m_{D}^{2},m_{K^{*}_{J}}^2,q^2)}}{\sqrt{q^2}}V_{0} \,.
\end{align}
\end{subequations}
with $\lambda(m_{D}^{2},m_{K^{*}_{J}}^2,q^2)=\left((m_{D}+m_{K^{*}_{J}})^2-q^2\right)\left((m_{D}-m_{K^{*}_{J}})^2-q^2\right)$.

The hadronic matrix element for $D$ decay to pseudoscalar meson $K^{*}_{0}(1430)$ can be written as
\begin{align}
\langle K^{*}_{1430}|\bar{s}\gamma^{\mu}\gamma_{5}c|D\rangle=(p_{D}+p_{K^{*}_{1430}})^{\mu}f_{+}(q^2)+q_{\mu}f_{-}(q^2) \,.
\end{align}
where $f_{+}$ and $f_{-}$ are the form factors of transition amplitudes for the charmed meson decaying to pseudoscalar $K^{*}_{1430}$.

Using the above transition, we can obtain the corresponding hadronic matrix elements
\begin{subequations}
\begin{align}
\tilde{H_{0}}=&\frac{\sqrt{\lambda(m_{D}^{2},m_{K^{*}_{1430}}^2,q^2)}}{\sqrt{q^2}}f_{+}(q^2) \,,\\
\tilde{H_{t}}=&\frac{f_{+}(q^2)(m_{D}^{2}-m_{K^{*}_{1430}}^2)+f_{-}(q^2)q^2}{\sqrt{q^2}} \,.
\end{align}
\end{subequations}

The decay amplitude of  $K^{*}_{J}\to K\pi$ for a vector meson can be  in terms of the couplings constant $g_{k^{*}}$, we obtain
\begin{align}
\langle K^{\pm}(p)\pi^{\mp}(p_{\pi})|K^{*}_{J}(p_{K^{*}_{J}},s_{K^{*}_{J}})\rangle=c_{\pm\mp}g_{k^{*}}(\epsilon_{s_{K^{*}_{J}}}\cdot p_{K^{*}_{J}}) \,.
\end{align}
The amplitude for a pseudoscalar meson $K^{*}_{1430}$ decaying into $K\pi$ is proportional to the $g_{k^{*}_{0}}$.  It is given by
\begin{align}
\langle K^{\pm}(p_{K})\pi^{\mp}(p_{\pi})|K^{*}_{1430}(p_{K^{*}_{J}},s_{K^{*}_{1430}})\rangle=c_{\pm\mp}g_{k^{*}_{0}} \,,
\end{align}
where $p_{K}$ and $s_{K^{*}_{J}}$ are the momentum of $K$ meson and the helicity of the $K^{*}$, respectively.  $c_{\pm\mp}$ represents the isospin factor of the K$\pi$ system. Here the subscript $\pm\mp$ represents the charged of K and $\pi$. In this paper, we choose $|c_{\pm\mp}|=1$. The coupling constant $g_{k^{*}}$ and $g_{k^{*}_{0}}$ can be determined from the $K^{*}_{J}$ two-body decay width, which read
\begin{subequations}
\begin{align}
g_{k^{*}}=&\sqrt{\frac{24\pi m_{K^{*}_{J}}^2\Gamma(K^{*}_{J}\to K\pi)}{|c_{\pm\mp}|^2|\vec{p}_{K}|^3}}~~~(J=892,~1410) \,,\\
g_{k^{*}_{0}}=&\sqrt{\frac{8\pi m_{K^{*}_{1430}}^2\Gamma(K^{*}_{1430}\to K\pi)}{|c_{\pm\mp}|^2|\vec{p}_{K}|}}~~~(J=1430) \,,
\label{kcoupling}
\end{align}
\end{subequations}
with $\vec{p}_{K}=\sqrt{\lambda(m_{K^{*}_{J}}^2,m_{K}^2,m_{\pi}^2)/2m_{K^{*}_{J}}}$. They encode importance of understanding of nonperturbative strong interactions.

For the vector meson $K^{*}_{J}$ resonances, such as J=892,1410, the Breit-Wigner function can be parameterized as 
\begin{align}
BW_{K^{*}_{J}}(p_{K^{*}_{J}}^2)=\frac{i}{p_{K^{*}_{J}}^2-m_{K^{*}_{J}}^2+im_{K^{*}_{J}}\Gamma_{m_{K^{*}_{J}}}} \,,
\end{align}
where $p_{K^{*}_{J}}^{2}=(p_{K}+p_{\pi})^2$ represents the invariant mass of the $K\pi$ pair. For the pesudoscalar resonance state $K^{*}_{1430}$, we also need to consider the contribution from $K^{*}_{0}(800)\equiv\kappa$.  $K^{*}_{0}(800)$ is a broad scalar meson. Therefore it can still contribute in the region where $p_{K^{*}}^2$ is close to the invariant mass square of $m_{K^{*}_{1430}}^2$. Finally the Breit-Wigner function of scalar meson can be parameterized as 
\begin{align}
BW_{scalar}(p_{K^{*}}^2)=-\frac{g_{\kappa}}{p_{K^{*}}^2-(m_{\kappa}-i\Gamma_{\kappa}/2)^2}+\frac{1}{p_{K^{*}}^2-(m_{K^{*}_{1430}}-i\Gamma_{K^{*}_{1430}}/2)^2} \,,
\label{lineshape}
\end{align}
Here $g_{\kappa}$ is a complex parameter $g_{\kappa}=|g_{\kappa}|e^{i \rm{arg}(g_{\pi})}$. Therefore $g_{\pi}\in [\frac{\pi}{2},\pi]$ is the phase~\cite{Becirevic:2012dp}. 

The differential decay width is defined as
\begin{align}
d\Gamma=d\Phi_{4}\times \frac{1}{2m_{D}}(2\pi)^4|\mathcal{M}(D\to K^{*}_{J}(K\pi)\ell\nu)| \,.
\end{align}
Here, the four-body phase-space $d\Phi_{4}$ is determined by
\begin{align}
d\Phi_{4}=\frac{\lambda(m_{D}^{2},m_{K^{*}_{J}}^2,q^2)q^{2}(1-\hat{m_{\ell}}^2)\vec{p}_{K}}{128\times(2\pi)^{10} m_{D}^2 \sqrt{p_{K^{*}_{J}}^2}}d\cos\theta_{K}d\cos\theta_{\ell}d\phi dq^2 dp_{K^{*}_{J}}^2 \,,
\end{align}
with $\hat{m_{\ell}}=m_{\ell}/\sqrt{q^2}$.

\subsection{Angular Distribution}
Taking into account effect from all vector and pseudoscalar resonance, the angular distribution for the four-body decay $D^{+}\to K^{*}_{892,1410,1430}(\to K^{\pm}\pi^{\mp})\ell^{+}\nu$ is then given by
\begin{align}\label{AD_A}
\frac{d\Gamma(D\to K^{*}_{J}(K\pi)\ell\nu)}{d\cos\theta_{K}d\cos\theta_{\ell}d\phi dq^2 dp_{K^{*}_{J}}^2}=&\frac{G_{F}^2 V_{cs}^2}{2}\frac{\lambda(m_{D}^{2},m_{K^{*}_{J}}^2,q^2)q^2(1-\hat{m_{\ell}}^2)^{2}\vec{p}_{K}}{4096\pi^{6}m_{D}^3\sqrt{p_{K^{*}_{J}}^2}}\nonumber\\
&\Big[ \big((\mathcal{A}_{11}\cos2\theta_{\ell}+\mathcal{A}_{12})\cos2\theta_{K}+\mathcal{A}_{13}\cos2\theta_{\ell}+\mathcal{A}_{14}\big)\cos2\phi\nonumber\\
&+\big((\mathcal{A}_{21}\sin2\theta_{\ell}+\mathcal{A}_{22}\sin\theta_{\ell})\sin2\theta_{K}+\mathcal{A}_{23}\sin2\theta_{\ell}+\mathcal{A}_{24}\sin\theta_{\ell}\big)\cos\phi \nonumber\\
&+((\mathcal{A}_{31}\cos2\theta_{\ell}+\mathcal{A}_{32})\cos2\theta_{K}+\mathcal{A}_{33}\cos2\theta_{\ell}+\mathcal{A}_{34})\sin2\phi\nonumber\\
&+((\mathcal{A}_{41}\sin2\theta_{\ell}+\mathcal{A}_{42}\sin\theta_{\ell})\sin2\theta_{K}+(\mathcal{A}_{43}\sin2\theta_{\ell}+\mathcal{A}_{44}\sin\theta_{\ell})\sin\theta_{K})\sin\phi\nonumber\\
&+\big((\mathcal{A}_{51}\cos2\theta_{\ell}+\mathcal{A}_{52}\cos\theta_{\ell}+\mathcal{A}_{53})\cos2\theta_{K}+(\mathcal{A}_{54}\cos2\theta_{\ell}+\mathcal{A}_{55}\cos\theta_{\ell})\cos\theta_{K}\nonumber\\
&+\mathcal{A}_{56}\cos2\theta_{\ell}+\mathcal{A}_{57}\cos\theta_{\ell}+\mathcal{A}_{58}\big) \Big] \,,
\end{align}
where the coefficients $A_{ij}(i=1\sim 5,j=1\sim 8)$ are listed in Appendix~\ref{apA}.

Employing the narrow-width approximation, we can safely neglect the interference term and only consider $K^{*}_{892}$ contribution. It is can be written 
\begin{align}
\int dp_{K^{*}}^{2}\frac{m_{K^{*}}\Gamma_{K^{*}_{J}}(K_{892}^{*}\to K \Pi)}{\pi}\frac{1}{(p_{K^{*}}^{2}-m_{K^{*}}^{2})^2+m_{K^{*}}^2\Gamma_{K^{*}}^2}=\mathcal{B}r(K_{892}^{*}\to K \pi) \,,
\end{align}
by integrating over the invariant mass of the $K\pi$ system, one can ascertain the threefold angular distribution that is characteristic of a P-wave in a four-body decay $D^{+}\to K^{*}_{892}(\to K\pi)\ell^{+}\nu$,
\begin{align}\label{AD_I}
\frac{d\Gamma(D\to K^{*}_{892}(\to K\pi)\ell\nu)}{d\cos\theta_{K}d\cos\theta_{\ell}d\phi dq^2 }=&\frac{3G_{F}^2 V_{cs}^2\lambda(m_{D}^{2},m_{K^{*}_{892}}^3,q^2)q^2(1-\hat{m_{\ell}}^2)^{2}\vec{p}_{K}\mathcal{B}r(K^{*}_{892}\to K\pi)}{128g_{K^{*}}^{2}\pi^{4}m_{D}^3((m_{K^{*}_{892}}-m_{K})^2-m_{\pi}^2)(m_{K^{*}_{892}}+m_{K})^2-m_{\pi}^2))^{3/2}}\nonumber\\
&\Big[ \big((\mathcal{I}_{11}\cos2\theta_{\ell}+\mathcal{I}_{12})\cos2\theta_{K}+\mathcal{I}_{13}\cos2\theta_{\ell}+\mathcal{I}_{14}\big)\cos2\phi\nonumber\\
&+\big((\mathcal{I}_{21}\sin2\theta_{\ell}+\mathcal{I}_{22}\sin\theta_{\ell})\sin2\theta_{K}\big)\cos\phi \nonumber\\
&+\big((\mathcal{I}_{31}\cos2\theta_{\ell}+\mathcal{I}_{32}\cos\theta_{\ell}+\mathcal{I}_{33})\cos2\theta_{K}\nonumber\\
&+\mathcal{I}_{34}\cos2\theta_{\ell}+\mathcal{I}_{35}\cos\theta_{\ell}+\mathcal{I}_{36}\big) \Big] \,.
\end{align}
Here the coefficients $\mathcal{I}_{ij}(i=1\sim 3,j=1\sim 6)$ are listed in Appendix~\ref{apA}. Due the $K^{*}_{1410}$ is also a P-wave states, the differentical decay width is similar to  Eq.\ref{AD_I}.

The scalar meson $K^{*}_{1430}$ four-body angular distribution also can be deduced by narrow-width approximation
\begin{align}
\frac{d\Gamma(D\to K^{*}_{1430}(\to K\pi)\ell\nu)}{d\cos\theta_{\ell}dq^2 }=&\frac{G_{F}^2 V_{cs}^2\lambda(m_{D}^{2},m_{K^{*}_{1430}}^3,q^2)q^2(1-\hat{m_{\ell}}^2)^{2}\vec{p}_{K}\mathcal{B}r(K^{*}_{1430}\to K\pi)}{268 g_{K_{0}^{*}}^{2}\pi^{3}m_{D}^3((m_{K^{*}_{1430}}-m_{K})^2-m_{\pi}^2)(m_{K^{*}_{1430}}+m_{K})^2-m_{\pi}^2))^{3/2}}\nonumber\\
&\Big[ d_{11}\cos2\theta_{\ell}+d_{12}\cos\theta_{l}+d_{13} \Big] \,,
\end{align}
with $d_{11}, d_{12}, d_{13}$ listed below
\begin{subequations}
\begin{align}
d_{11}=&(-1+\hat{m_{\ell}}^2)|\tilde{H}_{1430,0}|^2 \,,\\
d_{12}=&4\hat{m_{\ell}}^2\mathcal{R}e(\tilde{H}_{1430,0}^{*}\tilde{H}_{1430,t}) \,,\\
d_{13}=&2\hat{m_{\ell}}^2|\tilde{H}_{1430,t}|^2+(1+\hat{m_{\ell}}^2)|\tilde{H}_{1430,0}|^2 \,.
\end{align}
\end{subequations}
The angular distribution of the final state hadron in the two-body decays of a pseudoscalar is isotropic, which leads to the vanishment of $\theta_{k}$ and $\phi$ in four-body decays of angular distribution.

\section{Numerical Results}
\label{Panda}
\subsection{Inputs}
In this section, a comprehensive collection of meson parameters utilized in the numerical calculations is presented in Table~\ref{meson mass}. In addition, the CKM matrix element is given by $V_{cs}=0.975$. The masses of the electron and muon are $0.5\,\textrm{MeV}$ and $105\,\textrm{MeV}$, respectively  \cite{ParticleDataGroup:2022pth}. The branching fractions of $K^{*}$ are listed
\begin{align}
&\mathcal{B}(K^{*}_{892}\to K^{+}\pi^{-})=66.57\% \,,\qquad  \mathcal{B}(K^{*}_{892}\to K^{0}\pi^{0})=33.23\% \,,\nonumber \\
&\mathcal{B}(K^{*}_{1410}\to K^{+}\pi^{-})=4.67\% \,,\qquad  \mathcal{B}(K^{*}_{1410}\to K^{0}\pi^{0})=2.33\% \,,\nonumber \\
&\mathcal{B}(K^{*}_{1430}\to K^{+}\pi^{-})=66.67\% \,,\qquad \mathcal{B}(K^{*}_{1430}\to K^{0}\pi^{0})=33.33\% \,.\nonumber
\end{align}
For P-wave resonances $K^{*}_{892,1410}$, the decay coupling constant $g_{k^{*}}$ has been determined through lattice QCD calculations \cite{Prelovsek:2013ela}. Additionally, the S-wave decay coupling constant $g_{k_{0}^{*}}$ can be deduced from the decay width from Eq.\,(\ref{kcoupling}).

\begin{table}[!htbp]
  \centering
  \begin{tabular}{c|c|c|c|c|c}
  \hline
  \hline
  Mesons&$D^{\pm}$&$K^{*}_{892}$&$K^{*}_{1410}$&$K^{*}_{1430}$&$\kappa$  \\
  \hline
  $m$ (GeV)~&~1.896~&~0.891~&~1.414~&~1.425~&~0.658~\\
  \hline
  $\Gamma$ (GeV)~&~2.024~&~0.0514~ & ~0.232~ & ~0.27~ &~0.557~ \\
  \hline
  \end{tabular}
  \caption{The masses ($m$) and width ($\Gamma$) of concerned mesons in this work \cite{ParticleDataGroup:2022pth}.} \label{meson mass}
\end{table}
The form factors for the transition $D \to K^{*}$ have been calculated using the covariant light-front quark model framework and QCD sum rule methods. In the numerical analysis, one can obtain the form factors of P-wave mesons $K_{892,1410}^{*}$ is given by
\begin{align}
A/V_{i}(q^{2})=\frac{F(0)}{(1-q^{2}/m_{D}^2)(1-a_{i}(q^{2}/m_{D}^2)+b_{i}(q^{2}/m_{D}^2)^2)} \,.
\end{align}
As listed in Table.~\ref{ffp1}, $a_{i}$, $b_{i}$ and $F_{0}$ represent the fitting coefficients for different form factors $A$ and $V_{i=0,1,2}$~\cite{Cheng:2003sm,Ball:1991bs}.

\begin{table}[!htbp]
  \centering
  \begin{tabular}{c|c|c|c}
  \hline
  \hline
  Form Factors&$F_{0}$&a&b\\
  \hline
  A~&~0.94~&~1.17~&~0.42~\\
  \hline
 $V_{0}$&~0.69~&~1.04~ & ~0.44~  \\
   \hline
  $V_{1}$&~0.65~&~0.50~ & ~0.02~  \\
  \hline
   $V_{2}$&~0.57~&~0.94~ & ~0.27~  \\
  \hline
  \hline
  \end{tabular}
  \caption{Form factors for $D\to K^{*}$ transition obtained by covariant light-front quark model.} \label{ffp1}
\end{table}

 The pesudoscalar meson decay form factors can be expressed in the form
\begin{align}
f_{i}(q^{2})=\frac{F(0)}{1-a_{i}(q^{2}/m_{D}^2)+b_{i}(q^{2}/m_{D}^2)^2} \,,
\end{align}
where $a_{i}$ and $b_{i}$ ($i=+,-$) are fitting coefficients. They are listed in Table.~\ref{ffp2}

\begin{table}[!htbp]
  \centering
  \begin{tabular}{c|c|c|c}
  \hline
  \hline
  Form Factors&$F_{0}$&a&b\\
  \hline
  $f_{+}$&~0.48~&~1.01~&~0.24~\\
  \hline
 $f_{-}$&~0.48~&~-0.11~ & ~0.02~  \\
  \hline
  \hline
  \end{tabular}
  \caption{Form factors for $D\to K^{*}_{1430}$ transition obtained by covariant light-front quark model \cite{Cheng:2003sm}.} \label{ffp2}
\end{table}

\subsection{Numerical Results for $D^{+}\to K^{*}_{892,1410,1430}(\to K^{\pm}\pi^{\mp})\ell^{+}\nu$ }
By integrating over all angles, specifically  $\theta_{K}$, $\theta_{l}$ and $\phi$, one can obtain the decay width with three resonances,
\begin{align}
\frac{d\Gamma(D^{+}\to K^{*}_{J}(\to K^{\pm}\pi^{\mp})\ell^{+}\nu_{\ell})}{dp_{K^{*}}^2 dq^2}=\mathcal{P}\frac{8}{9}\pi \Big[ \mathcal{A}_{51}-3(\mathcal{A}_{53}+\mathcal{A}_{56}-3\mathcal{A}_{58}) \Big] \,,
\label{decay function}
\end{align}
with $\mathcal{P}=\frac{G_{F}^2 V_{cs}^2}{2}\frac{\lambda(m_{D}^{2},m_{K^{*}_{J}}^2,q^2)q^2(1-\hat{m_{\ell}}^2)^{2}\vec{p}_{K}}{4096\pi^{6}m_{D}^3\sqrt{p_{K^{*}_{J}}^2}}$ and $J=892,1410,1430$.  After setting the integration limits for $p_{K^{*}_{J}}^2$, the corresponding maximum value of $q^2$, expressed as $q^2=\left(m_{D}-\sqrt{p_{K^{*}_{J}}^2}\right)^2$, is determined to be dependent on $p_{K^{*}_{J}}^2$.  Ensuring adherence to kinematic constraints for all particles, we utilize the Heaviside function $\Theta$  within the differential decay width function,
\begin{align}
\frac{d\Gamma(D^{+}\to K^{*}_{J}(\to K^{\pm}\pi^{\mp})\ell^{+}\nu_{\ell})}{dp_{K^{*}}^2 dq^2}=\mathcal{P}\frac{8}{9}\pi \Big[\mathcal{A}_{51}-3(\mathcal{A}_{53}+\mathcal{A}_{56}-3\mathcal{A}_{58}) \Big]\Theta\left(m_{D}-\sqrt{p_{K^{*}_{J}}}-\sqrt{q^2}\right)\Theta\left(m_{D}-m_{K}-m_{\pi}\right) \,.
\end{align}

\begin{figure}[htbp!]
  \begin{minipage}[t]{0.4\linewidth}
  \centering
  \includegraphics[width=1.0\columnwidth]{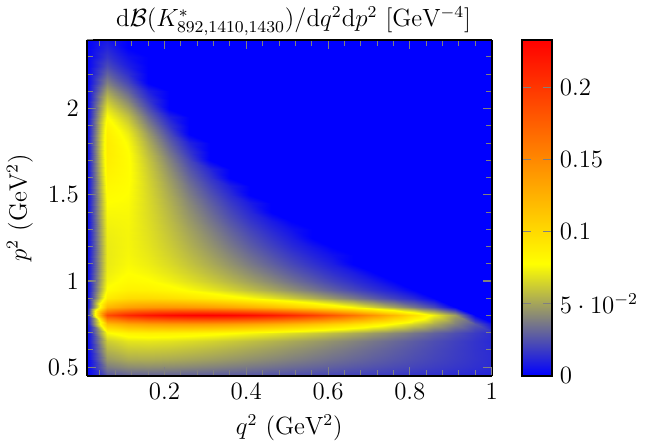}
    \end{minipage}
    \begin{minipage}[t]{0.4\linewidth}
  \centering
  \includegraphics[width=1.0\columnwidth]{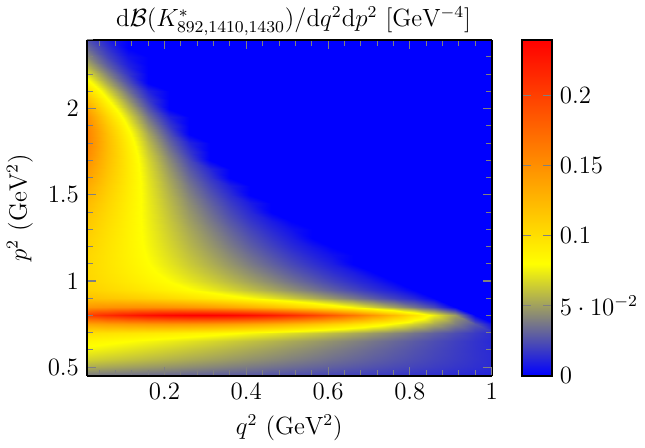}
    \end{minipage}
 \caption{ The differential decay branching fractions for $D^+\to K^*_J(\to K^{-}\pi^{+})\ell^+\nu_{\ell}$ with respect to $q^2$ and $p_{K^*_J}^2$ correspond to final states with muons (left) and electrons (right), respectively.} \label{decay width}
\end{figure}

In Figure \ref{decay width}, we present the numerical results for the differential distribution of the branching fraction with three resonances as a function of the invariant mass of the
$K\pi$ system $p_{K^{*}_{J}}^{2}$ and the squared momentum transfer $q^2$. It exhibits a peak around at $p_{K^{*}}^2\sim 0.8~ \rm{GeV}^{2}$ indicating a predominant contribution from $K^{*}_{892}$. An additional  bump is discernible in the branching fraction distributions, which corresponds to the contribution from $K^{*}_{1410}$ and $K^{*}_{1430}$ resonances. In contrast to the distinct resonances of $K^{*}_{1410}$ or $K^{*}_{1430}$, the observed bump in Figure~\ref{decay width} exhibits a broader width, which is attributed to the fact of that the mass of $K^{*}_{1410}$ is very close $K^{*}_{1430}$. 

\begin{figure}[htbp!]
  \begin{minipage}[t]{0.4\linewidth}
  \centering
  \includegraphics[width=1.0\columnwidth]{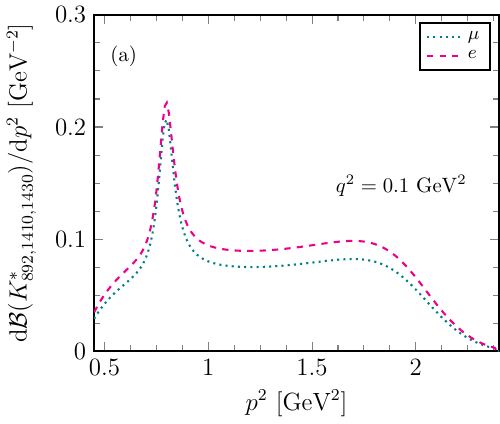}
    \end{minipage}
    \begin{minipage}[t]{0.4\linewidth}
  \centering
  \includegraphics[width=1.0\columnwidth]{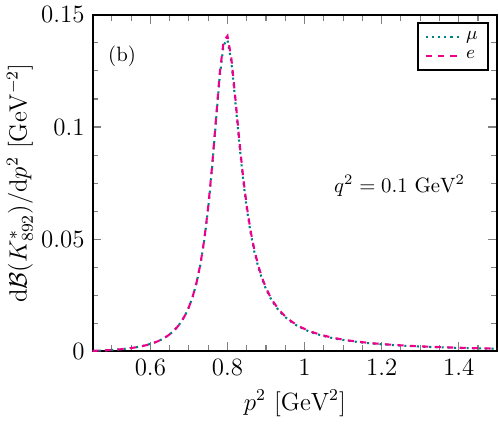}
    \end{minipage}
      \begin{minipage}[t]{0.4\linewidth}
  \centering
  \includegraphics[width=1.0\columnwidth]{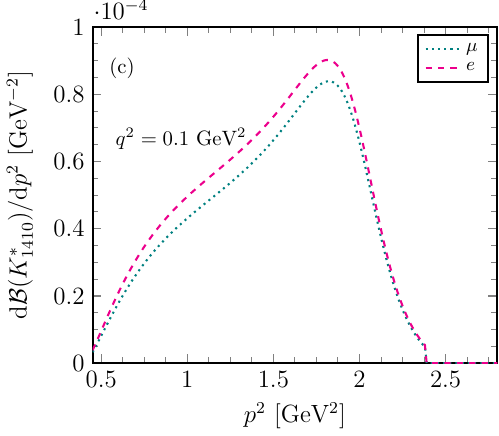}
    \end{minipage}
    \begin{minipage}[t]{0.4\linewidth}
  \centering
  \includegraphics[width=1.0\columnwidth]{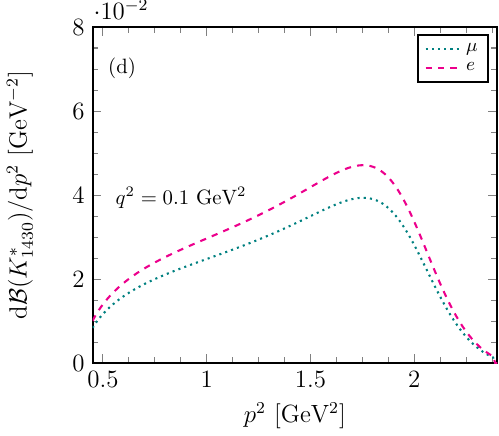}
    \end{minipage}
 \caption{ The differential decay branching fractions for $D^+\to K^*_{892,1410,1430}(\to K^{-}\pi^{+})\ell^+\nu_{\ell}$, $D^+\to K^*_{892}(\to K^{-}\pi^{+})\ell^+\nu_{\ell}$, $D^+\to K^*_{1410}(\to K^{-}\pi^{+})\ell^+\nu_{\ell}$, and $D^+\to K^*_{1430}(\to K^{-}\pi^{+})\ell^+\nu_{\ell}$ with $q^2=0.1 \rm~{GeV}^2$. } \label{Br resonance}
\end{figure}

To have a more clear picture of the above discussion, we present the differential decay branching fractions at a typical value of $q^2 = 0.1\,\text{GeV}^2$ in Fig.\ref{Br resonance}. A narrow width peak is observed at approximately $0.8\,\text{GeV}^2$, which is attributed to the decay process $D \to K^*_{892}$. Additionally, It highlights that the $K^{*}_{1430}$ resonance bump shown in Figure \ref{Br resonance} (d) has a notably greater influence on the branching fraction than the $K^{*}_{1410}$. This is particularly evident in the pronounced flat shoulder observed within the energy range from $1\,\rm{GeV}^2$ to $2\,\rm{GeV}^2$, which is caused by a broad width peak from $K^{*}_{1430}$ and the contribution from $K^{*}_{892}$ tending to zero. Furthermore, the calculations for the $D^+\to K^*_{892,1410,1430}(\to K^{-}\pi^{+})\ell^+\nu_{\ell}$ decay distribution include the effects of interference among the three contributing resonances. Compared to the  $D \to K^*_{892}$ decay result, the distribution exhibits larger contributions, which are primarily attributed to the presence of the additional $K^{*}_{1430}$ and the positive correction from the interference term.

\begin{figure}[htbp!]
  \begin{minipage}[t]{0.4\linewidth}
  \centering
  \includegraphics[width=1.0\columnwidth]{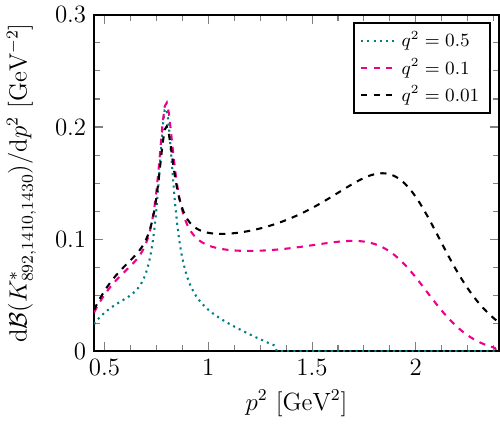}
    \end{minipage}
 \caption{The differential distributions of  $K^{*}_{892,1410,1430}$  with $q^2=0.5\rm{GeV}^2$ (doted green line), $q^2=0.1\rm{GeV}^2$  (red dashed line) and $q^2=0.01\rm{GeV}^2$  (black dashed line). } \label{smallq2}
\end{figure}

Furthermore, at higher values of $q^2$, the contributions from $K^*_{892,1410,1430}$  resonances are  significantly suppressed and eventually we can only find a single peak structure. On the other hand, at low $q^2$, the phase-phase for $K^{*}_{1430}$ production has been fully accessible, as illustrated in Fig.\ref{smallq2}.

\begin{figure}[htbp!]
  \begin{minipage}[t]{0.4\linewidth}
  \centering
  \includegraphics[width=1.0\columnwidth]{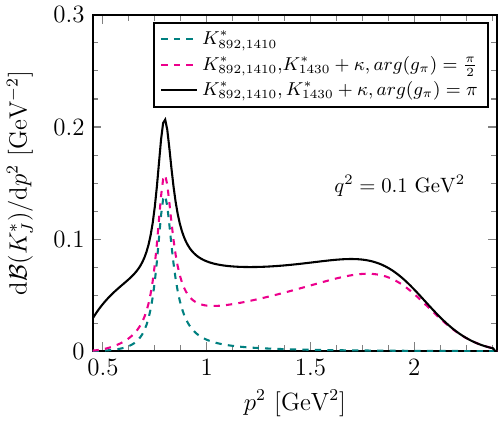}
    \end{minipage}
 \caption{Line shapes of the resonant $K^{*}_{892,1410}$ (dashed green line),  $K^{*}_{892,1410,1430, \kappa}$ with $\rm{arg}$$(g_{\kappa})=\pi/2$ (red dashed line), $\rm{arg}$$(g_{\kappa})=\pi$ (black solid line).} \label{delta}
\end{figure}

The mass and width of the scalar meson $\kappa$ are taken from ref \cite{Descotes-Genon:2006sdr}. The magnitude and argument ranges of the coupling $g_{\kappa}$, as depicted in Eq.\,(\ref{lineshape}), are detailed in the ref~\cite{Becirevic:2012dp}. In this section, we use a specific value for the coupling constant  $|g_{\kappa}|=0.2$. Additionally, we choose the phase of $\rm{arg}(g_{\pi})$, varying it from $\pi/2$ to $\pi$ as in~\cite{Das:2017ebx}. In the following, we investigate the impact of the phase. To illustrate the contribution of scale $\rm{arg}(g_{\pi})$, we first display the behavior of $K^{*}_{892,1410}$ distribution in Fig \ref{delta}.
Then, we present the distributions of the P-wave $K^{*}_{892,1410}$  and pseudoscalar mesons $K^{*}_{1430}$ and $\kappa$. Both of the lines in Fig. \ref{delta} indicate the $K^{*}_{892}$ exhibits a pronounced peak in its P-wave contribution, consistent with the conditions detailed in Fig. \ref{Br resonance}(a). Conversely, the P-wave contribution of the  $K^{*}_{1410}$ decreases towards zero with increasing squared momentum $p^{2}$, attributed to its minimal contribution. Furthermore, scalar mesons exhibit a positive contribution at higher squared momentum $p^{2}$  values. As depicted in Fig. \ref{delta}, varying the coupling constant $g_{\kappa}$ results in either a flat shoulder or a bump distribution, depending on its value. When the argument $\rm{arg}(g_{\pi})$ approaches $\pi/2$, a bump emerges in the vicinity of the invariant mass square of the $K^*_{1430}$. As the argument of $\rm{arg}(g_{\pi})$ continues to increase,  the scalar mesons exert a positive influence on the higher invariant mass ranges within the $K\pi$ system, leading to a flat shoulder. So the shape of the differential distributions  can help experiments to determine the value of $\rm{arg}(g_{\pi})$. 

\subsection{Numerical Results for $D^{+}\to K^{*}_{892}(\to K^{-}\pi^{+})\ell^{+}\nu$ }
\begin{figure}[htbp!]
  \begin{minipage}[t]{0.4\linewidth}
  \centering
  \includegraphics[width=1.0\columnwidth]{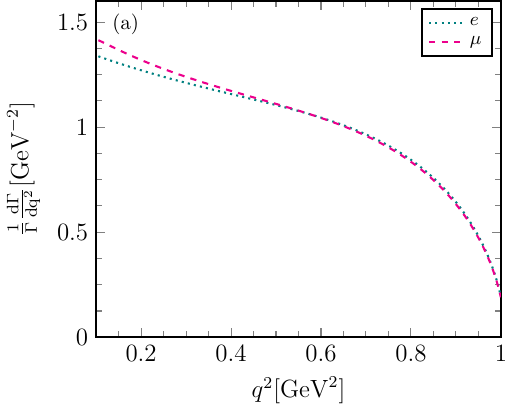}
    \end{minipage}
    \begin{minipage}[t]{0.4\linewidth}
  \centering
  \includegraphics[width=1.0\columnwidth]{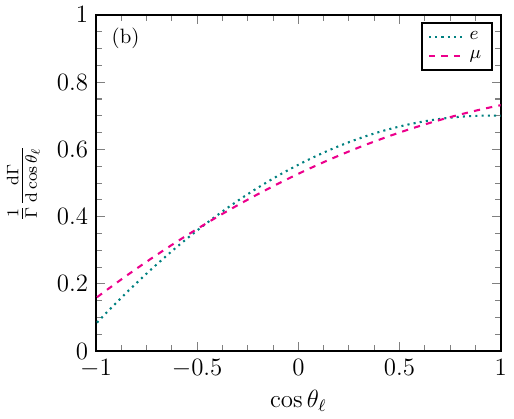}
    \end{minipage}
      \begin{minipage}[t]{0.4\linewidth}
  \centering
  \includegraphics[width=1.0\columnwidth]{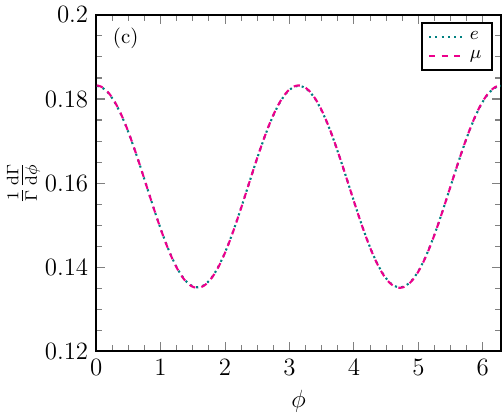}
    \end{minipage}
    \begin{minipage}[t]{0.4\linewidth}
  \centering
  \includegraphics[width=1.0\columnwidth]{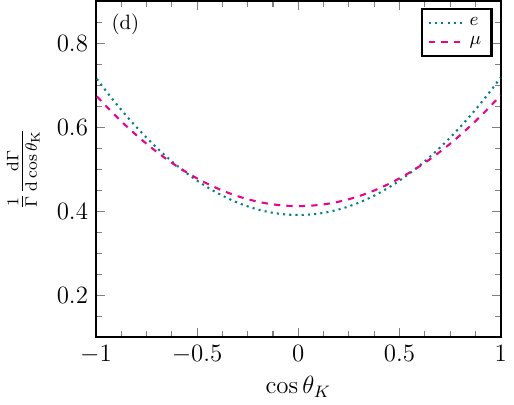}
    \end{minipage}
 \caption{ The differential decay branching fractions for $D^+\to K^*_{892}(\to K^{-}\pi^{+})\ell^+\nu_{\ell}$.} \label{angular}
\end{figure}

In our numerical analysis, we concentrate on the predominant excited states of the $K^{*}_{892}$. By integrating the angles $\theta_{l}$, $\theta_{K}$ and $\phi$ from the differential decay width, we deduce the normalized distribution for $q^2$ which is shown in Figure \ref{angular}(a).

The normalized distribution of $\cos\theta_{l}$
\begin{align}
\frac{1}{\Gamma}\frac{d\Gamma(D\to K^{*}_{892}(\to K^{-}\pi^{+})\ell\nu)}{d\cos\theta_{\ell} dq^2}=-\frac{\mathcal{P}}{\Gamma}\frac{4}{3}\pi(L_{\theta_{\ell}0}+L_{\theta_{\ell}1}\cos\theta_{\ell}+L_{\theta_{\ell}2}\cos2\theta_{\ell}) \,,
\end{align}
where $L_{\theta_{\ell}0}=\mathcal{I}_{53}-3\mathcal{I}_{56}$, $L_{\theta_{\ell}1}=\mathcal{I}_{52}-3\mathcal{I}_{55}$ and $L_{\theta_{\ell}2}=\mathcal{I}_{51}-3\mathcal{I}_{54}$.

The normalized distribution of $\phi$
\begin{align}
\frac{1}{\Gamma}\frac{d\Gamma(D\to K^{*}_{892}(\to K^{-}\pi^{+})\ell\nu)}{d\phi dq^2}=\frac{\mathcal{P}}{\Gamma}\frac{4}{9}(L_{\phi 0}+L_{\phi 1}\cos2\phi) \,,
\label{thetak}
\end{align}
where $L_{\phi 0}=\mathcal{I}_{51}-3\mathcal{I}_{53}-3\mathcal{I}_{54}+9\mathcal{I}_{56}$ and $L_{\phi 1}=\mathcal{I}_{11}-3(\mathcal{I}_{12}-3\mathcal{I}_{13}+9\mathcal{I}_{14})$.

The normalized distribution of $\cos\theta_{k}$
\begin{align}
\frac{1}{\Gamma}\frac{d\Gamma(D\to K^{*}_{892}(\to K^{-}\pi^{+})\ell\nu)}{d\cos\theta_{K} dq^2}=-\frac{\mathcal{P}}{\Gamma}\frac{4}{3}\pi(L_{\theta_{K}0}+L_{\theta_{k}2}\cos2\theta_{K}) \,,
\end{align}
where $L_{\theta_{K}0}=\mathcal{I}_{54}-3\mathcal{I}_{56}$ and $L_{\theta_{K}2}=\mathcal{I}_{51}-3\mathcal{I}_{53}$.

The distributions of  $\cos\theta_{\ell}$, $\phi$ and $\cos\theta_{K}$ are depicted in Figure \ref{angular}(b,c,d). The absence of a coefficient for $\cos\theta_{K}$  in Eq. (\ref{thetak}) indicates that there is no dependence on the angle $\theta_{K}$,  which implies that forward-backward asymmetry is not present. Consequently, this suggests that the unpolarized intermediate states of vector mesons decaying into two pseudoscalars would not exhibit Forward-Backward asymmetry. The intermediate state, as a scalar meson decaying into two pseudoscalar mesons, exhibits isotropic behavior, leading to no dependence on the angle $\theta_{K}$. Hence, the decays of D mesons through intermediate states such as $K^{*}(892)$, $K^{*}(1410)$ and $K_{0}^{*}(1430)$ do not exhibit forward-backward asymmetry.

 The angular distributions distinctly elucidate the fundamental universality characteristic of leptonic decays. Upon integrating the complete set of parameters, we derive the branching fractions:  $\mathcal{B}r(D^{+}\to K^{*}_{892}(\to K^{-}\pi^{+})e^{+}\nu_{e})=3.61\%$ and $\mathcal{B}r(D^{+}\to K^{*}_{892}(\to K^{-}\pi^{+})\mu^{+}\nu_{\mu})=3.25\%$. Therefore, the ratio of the branching fractions obtained is
\begin{align}
  \frac{\mathcal{B}r(D^{+}\to K^{*}_{892}(K^{-}\pi^{+}) \, \mu^{+}\nu_{\mu})}{\mathcal{B}r(D^{+}\to K^{*}_{892}(K^{-}\pi^{+})  \, e^{+}\nu_{e})} = 0.975 \,.
\end{align}
This result is consistent with Particle Data Group~\cite{ParticleDataGroup:2022pth} and  in agreement with LFU. Based on our detailed analysis of the $K_{892}^{*}$ resonance, we can project the branching fractions for the decays $\mathcal{B}r(D^{+}\to K^{*}_{1410}(\to K^{-}\pi^{+})\ell^{+}\nu_{\ell})$ and $\mathcal{B}r(D^{+}\to K^{*}_{1430}(\to K^{-}\pi^{+})\ell^{+}\nu_{\ell})$
\begin{align}
\mathcal{B}r(D^{+}\to K^{*}_{1410}(\to K^{-}\pi^{+})e^{+}\nu_{e})=0.007\% \,,\\ \nonumber
 \mathcal{B}r(D^{+}\to K^{*}_{1410}(\to K^{-}\pi^{+})\mu^{+}\nu_{\mu})=0.005\% \,,\\ \nonumber
 \mathcal{B}r(D^{+}\to K^{*}_{1430}(\to K^{-}\pi^{+})e^{+}\nu_{e})=0.053\% \,,\\ \nonumber
 \mathcal{B}r(D^{+}\to K^{*}_{1430}(\to K^{-}\pi^{+})\mu^{+}\nu_{\mu})=0.041 \%\,.
\end{align}
Consequently, the relative proportions of the branching fractions are determined as
\begin{equation}
\begin{split}
  \frac{\mathcal{B}r(D^{+}\to K^{*}_{1410}(K^{-}\pi^{+}) \, \mu^{+}\nu_{\mu})}{\mathcal{B}r(D^{+}\to K^{*}_{1410}(K^{-}\pi^{+})  \, e^{+}\nu_{e})} =& 0.714  \,,\\
   \frac{\mathcal{B}r(D^{+}\to K^{*}_{1430}(K^{-}\pi^{+}) \, \mu^{+}\nu_{\mu})}{\mathcal{B}r(D^{+}\to K^{*}_{1430}(K^{-}\pi^{+})  \, e^{+}\nu_{e})} =&0.774 \,.
  \end{split}
\end{equation}

The ratios of $K^{*}_{1410}$ and $K^{*}_{1430}$ are significantly different from that of  $K^{*}_{892}$. As the mass of the intermediate states increases, it leads to a compression of the  existence space of the $q^2$. Consequently, the mass of the final states significantly influences the branching fractions.


\section{Conclusions}
\label{Conclusions}
Weak decays of heavy quarks have played an important role in testing SM and probing NP. Research on the semileptonic $D$ decays is of significant importance in calibrating relevant nonleptonic $D$ decays and semileptonic $B$ decays, investigating the properties of the $K^*_J$ resonance, as well as testing the LFU. We have in this work systematically derived differential decay widths and angular distributions for the decay cascade $D^{+}\to  K^{*}_{J} \ell^{+}\nu \to  K^{\pm/0}\pi^{\mp/0}  \ell^{+}\nu$. In the derivation, the mass of electron and muon is explicitly included. Our results are consistent with the existing experimental data and demonstrate the flexibility of the form factors proposed in \cite{Cheng:2003sm}. Besides, we derived the ratios of branching fractions $\frac{\mathcal{B}r(D^{+}\to K^{*}(892)(K^{-}\pi^{+}) \, \mu^{+}\nu_{\mu})}{\mathcal{B}r(D^{+}\to K^{*}(892)(K^{-}\pi^{+}) \, e^{+}\nu_{e})} = 0.975$, $\frac{\mathcal{B}r(D^{+}\to K^{*}(1410)(K^{-}\pi^{+}) \, \mu^{+}\nu_{\mu})}{\mathcal{B}r(D^{+}\to K^{*}(1410)(K^{-}\pi^{+}) \, e^{+}\nu_{e})} = 0.714$ and $\frac{\mathcal{B}r(D^{+}\to K^{*}_{0}(1430)(K^{-}\pi^{+}) \, \mu^{+}\nu_{\mu})}{\mathcal{B}r(D^{+}\to K^{*}_{0}(1430)(K^{-}\pi^{+}) \, e^{+}\nu_{e})} = 0.774$. We hope these studies will provide a better understanding of the semileptonic decay dynamics.

\section*{Acknowledgements}
We thank Prof. Lei Li, Wei Wang and Yateng Zhang for fruitful discussions on the semileptonic decays of $D$ meson. This work is supported in part by Natural Science Foundation of China under grant No. 12105247.

\begin{appendix}
\section{Coefficient function in angular distribution}\label{apA}

The coefficients $A_{ij}$ in Eq.\,(\ref{AD_A}) are presented below:
\begin{align}
\mathcal{A}_{11}=&\mathcal{A}_{14}=-\mathcal{A}_{12}=-\mathcal{A}_{13}=\frac{1}{32}(-1+\hat{m_{\ell}}^2)\big(\sum_{J}H_{-1,J}^{*}\sum_{J'}H_{1,J'}+h.c.\big)\;,\\ \nonumber
\mathcal{A}_{21}=&\frac{1}{16}(-1+\hat{m_{\ell}}^2)\big((\sum_{J^{}}H_{0,J^{}}(\sum_{J}H_{-1,J}^{*}+\sum_{J}H_{1,J'}^{*})+h.c.))\;,\\ \nonumber
\mathcal{A}_{22}=&\frac{-1}{8}\big(\hat{m_{\ell}^{2}}(\sum_{J}H_{t,J}(\sum_{J'}H_{-1,J'}^{*}+\sum_{J'}H_{1,J'})+\sum_{J}H_{0,J}^{*}(\sum_{J'}H_{-1,J'}-\sum_{J'}H_{1,J'})+h.c.))\;,\\ \nonumber
\mathcal{A}_{23}=&\frac{1}{4}(-1+\hat{m_{\ell}^2})(H_{0,1430}(\sum_{J}H_{-1,J}^{*}+\sum_{J}H_{1,J}^{*})+h.c.))\;, \\ \nonumber
\mathcal{A}_{24}=&\frac{1}{2}(H_{0,1430}^{*}(\sum_{J}H_{1,J}-\sum_{J}H_{-1,J})+\hat{m_{\ell}^2}(H_{t,1430}(\sum_{J}H_{-1,J}^{*}+\sum_{J}H_{1,J}^{*}))+h.c.)\;, \\ \nonumber
\mathcal{A}_{31}=&\mathcal{A}_{34}=-\mathcal{A}_{32}=-\mathcal{A}_{33}=\frac{1}{32}(-1+\hat{m_{\ell}}^2)\mathcal{I}m\big(\sum_{J}H_{-1,J}\sum_{J'}H_{1,J'}^{*}-h.c.\big)\\ \nonumber
\mathcal{A}_{41}=&\frac{1}{16}(-1+\hat{m_{\ell}}^2)\mathcal{I}m\big(\sum_{J'}H_{0,J'}(\sum_{J}H_{-1,J}^{*}-\sum_{J}H_{1,J}^{*})+h.c.\big)\\ \nonumber
\mathcal{A}_{42}=&\frac{1}{8}\big(\hat{m_{\ell}}^2\mathcal{I}m(\sum_{J}H_{t,J}(\sum_{J'}H_{1,J'}^{*}-\sum_{J'}H_{-1,J'}^{*}+h.c.)+\sum_{J}H_{0,J}^{*}(\sum_{J'}H_{-1,J'}+\sum_{J'}H_{1,J'})-h.c.)\big)\\ \nonumber
\mathcal{A}_{43}=&\frac{1}{4}(-1+\hat{m_{\ell}^2})\mathcal{I}m(H_{0,1430}^{*}(\sum_{J}H_{-1,J}^{*}-\sum_{J}H_{1,J}^{*})+h.c.) \\ \nonumber
\mathcal{A}_{44}=&\frac{1}{2}\mathcal{I}m(-H_{0,1430}^{*}(\sum_{J}H_{-1,J}+\sum_{J}H_{1,J})+h.c.)+\hat{m_{\ell}^2}(H_{t,1430}^{*}(\sum_{J}H_{-1,J}-\sum_{J}H_{1,J})+h.c.) \\ \nonumber
\mathcal{A}_{51}=&\frac{1}{32}(-1+\hat{m_{\ell}}^2)\big( \sum_{J}H_{-1,J}^{*}\sum_{J}H_{-1,J}+\sum_{J}H_{1,J}^{*}\sum_{J}H_{1,J}+4\sum_{J}H_{0,J}^{*}\sum_{J}H_{0,J}\big)\;,\\ \nonumber
\mathcal{A}_{52}=&\frac{1}{8}\big(\sum_{J}H_{-1,J}^{*}\sum_{J}H_{-1,J}+\sum_{J}H_{1,J}^{*}\sum_{J}H_{1,J}+2\hat{m_{\ell}^2}(\sum_{J}H_{t,J}^{*}\sum_{J}H_{0,J}+\sum_{J}H_{0,J}^{*}\sum_{J}H_{t,J})\big)\;,\\ \nonumber
\mathcal{A}_{53}=&\frac{1}{32}(-(3+\hat{m_{\ell}}^2)\sum_{J}H_{-1,J}^{*}\sum_{J}H_{-1,J}-4(1+\hat{m_{\ell}}^2)\sum_{J}H_{0,J}^{*}\sum_{J}H_{0,J}+\\ \nonumber
&(3+\hat{m_{\ell}}^2)\sum_{J}H_{1,J}^{*}\sum_{J}H_{1,J}-8\hat{m_{\ell}}^2\sum_{J}H_{t,J}^{*}\sum_{J}H_{t,J}))\;,\\ \nonumber
\mathcal{A}_{54}=&\frac{-1}{2}(-1+\hat{m_{\ell}}^2)\big(H_{0,1430}^{*}\sum_{J}H_{0,J}+h.c.\big)\;,\\ \nonumber
\mathcal{A}_{55}=&-\hat{m_{\ell}}^2(H_{t,1430}\sum_{J}H_{0,J}^{*}+\sum_{J}H_{0,1430}\sum_{J}H_{t,J}^{*}+h.c.)\\ \nonumber
\mathcal{A}_{56}=&\frac{-1}{32}\big((-1+\hat{m_{\ell}}^2)\sum_{J}H_{-1,J}^{*}\sum_{J}H_{-1,J} +4\sum_{J}H_{0,J}^{*}\sum_{J}H_{0,J}+\\ \nonumber
&(-1+\hat{m_{\ell}}^2)\sum_{J}H_{1,J}^{*}\sum_{J}H_{1,J}+(-1+\hat{m_{\ell}}^2)H_{0,1430}^{*}H_{0,1430}\big)\;,\\ \nonumber
\mathcal{A}_{57}=&\frac{1}{8}\big( 16\hat{m_{\ell}}^2(H_{0,1430}^{*}H_{t,1430}+H_{0,1430}H_{t,1430}^{*})+(\sum_{J}H_{-1,J}^{*}\sum_{J}H_{-1,J}-\sum_{J}H_{1,J}^{*}\sum_{J}H_{1,J})+\\ \nonumber
&2\hat{m_{\ell}}^2(\sum_{J}H_{0,J}^{*}\sum_{J}H_{t,J}+\sum_{J}H_{0,J}\sum_{J}H_{t,J}^{*})\big)\;,\\ \nonumber
\mathcal{A}_{58}=&\frac{1}{32}\hat{m_{\ell}}^2\big( 32(H_{t,1430}^{*}H_{t,1430}+H_{0,1430}H_{0,1430}^{*})+(\sum_{J}H_{-1,J}^{*}\sum_{J}H_{-1,J}+\sum_{J}H_{1,J}^{*}\sum_{J}H_{1,J})+4\sum_{J}H_{0,J}^{*}\sum_{J}H_{0,J}\\ \nonumber
&+8\sum_{J}H_{t,J}^{*}\sum_{J}H_{t,J}+\frac{1}{32}(H_{0,1430}H_{0,1430}^{*}+3(\sum_{J}H_{-1,J}^{*}\sum_{J}H_{-1,J}+\sum_{J}H_{1,J}^{*}\sum_{J}H_{1,J})+4\sum_{J}H_{0,J}^{*}\sum_{J}H_{0,J})\big).\\ \nonumber
\end{align}
Here $J$ symbolizes the resonances associated with the particles denoted as $K^{*}_{892}$, and $K^{*}_{1410}$.

The coefficients $\mathcal{I}_{ij}$ in Eq.\,(\ref{AD_I}) are listed:
\begin{align}
\mathcal{I}_{11}=&\mathcal{I}_{14}=-\mathcal{I}_{12}=-\mathcal{I}_{13}=\frac{1}{16}(-1+\hat{m_{\ell}}^2)\mathcal{R}e(H_{-1,892}^{*}H_{1,892})\;,\\ \nonumber
\mathcal{I}_{21}=&-\frac{1}{8}(-1+\hat{m_{\ell}}^2)\mathcal{R}e(H_{0,892}H_{-1,892}^{*})\;,\\ \nonumber
\mathcal{I}_{22}=&\frac{1}{8}\big(-2\hat{m_{\ell}^{2}}\mathcal{R}e(H_{t,892}^{*}(H_{-1,892}+H_{1,892})+H_{0,892}^{*}(H_{-1,892}-H_{1,892})))\;,\\ \nonumber
\mathcal{I}_{31}=&\frac{1}{32}(-1+\hat{m_{\ell}^2})(|H_{-1,892}|^2+4|H_{0,892}|^2+|H_{1,892}|^2)\;, \\ \nonumber
\mathcal{I}_{32}=&\frac{1}{8}(-|H_{-1,892}|^2+4\hat{m_{\ell}}^2\mathcal{R}e(H_{0,892}H_{t,892}^{*})+|H_{1,892}|^2)\;, \\ \nonumber
\mathcal{I}_{33}=&-\frac{1}{32}((3+\hat{m_{\ell}}^2)|H_{-1,892}|^2-4(1+\hat{m_{\ell}}^2)|H_{0,892}|^2+(3+\hat{m_{\ell}}^2)|H_{1,892}|^2-8\hat{m_{\ell}}^2|H_{t,892}|^2)\;, \\ \nonumber
\mathcal{I}_{34}=&-\frac{1}{32}(-1+\hat{m_{\ell}^2})(|H_{-1,892}|^2+|H_{0,892}|^2+|H_{1,892}|^2)\;, \\ \nonumber
\mathcal{I}_{35}=&\frac{1}{8}(|H_{-1,892}|^2+4\hat{m_{\ell}}^2\mathcal{R}e(H_{0,892}H_{t,892}^{*})-|H_{1,892}|^2)\;, \\ \nonumber
\mathcal{I}_{36}=&\frac{1}{32}((3+\hat{m_{\ell}}^2)|H_{-1,892}|^2+4(1+\hat{m_{\ell}}^2)|H_{0,892}|^2+(3+\hat{m_{\ell}}^2)|H_{1,892}|^2+8\hat{m_{\ell}}^2|H_{t,892}|^2)\;.\\ \nonumber
\end{align}

\end{appendix}

\end{document}